\documentclass[10pt,a4paper]{article}
\usepackage{fullpage}
\usepackage[latin1] {inputenc}
\usepackage{amsmath}
\usepackage{amsfonts}
\usepackage{amssymb}
\usepackage{enumerate}
\usepackage{url}

\newtheorem{definition}{Definition}[section]

\newtheorem{lemma}[definition]{Lemma}

\newtheorem{theorem}[definition]{Theorem}
\newtheorem{cor}[definition]{Corollary}
\newtheorem{obs}[definition]{Observation}

\newtheorem{claim}{Claim}

\newcommand{\Prob}[1]{\mathbf{P} \left( #1 \right)}
\newcommand{\Expec}[1]{\mathbf{E} \left[ #1 \right]}
\newcommand{\proof}{\noindent\textit{Proof. }}
\newcommand{\qed}{\hspace{\stretch{1}$\square$}}

\newcommand{\polylog }{\mbox{polylog }}

\begin{document}

\title{\bf{Information Spreading in \\ Stationary Markovian Evolving Graphs}\thanks{A preliminary version of this work was presented at the \emph{24th IEEE IPDPS} 2009}}
\author{Andrea~Clementi\thanks{Dipartimento di Matematica, Universit\`a di  Tor Vergata, Rome, Italy, \texttt{clementi@mat.uniroma2.it}}
\and
Angelo~Monti\thanks{Dipartimento di Informatica, Sapienza Universit\`a di Roma, Rome, Italy, \texttt{monti/silvestri@di.uniroma1.it}}
\and
Francesco~Pasquale\thanks{Dipartimento di Informatica e Applicazioni ``Renato M. Capocelli'', Universit\`a di Salerno, Salerno, Italy, \texttt{pasquale@dia.unisa.it}}
\and
Riccardo~Silvestri$^\ddagger$}
\date{}

\maketitle

\begin{abstract}
\emph{Markovian evolving graphs}  are dynamic-graph models where the links among a fixed set of nodes change during time according to an arbitrary Markovian rule. They are extremely general and they can well describe important dynamic-network scenarios.

We study the speed of information spreading in the \emph{stationary phase} by analyzing the completion time of the \emph{flooding mechanism}. We prove a general theorem that establishes an upper bound on flooding time in \emph{any} stationary Markovian evolving graph in terms of its node-expansion properties.

We apply our theorem in two natural and relevant cases of such dynamic graphs. \emph{Geometric Markovian evolving graphs}  where the Markovian behaviour is yielded by $n$ mobile radio stations, with fixed transmission radius, that perform   independent random walks over a square region of the plane.  \emph{Edge-Markovian evolving graphs}  where the probability of existence of any edge at time $t$ depends on the existence (or not) of the same edge at time $t-1$.

In both cases, the obtained upper bounds hold \emph{with high probability} and they are nearly tight. In fact, they turn out to be tight for a large range of the values of the input parameters. As for geometric Markovian evolving graphs, our result represents the first analytical upper  bound for flooding time on a class of concrete mobile networks.

\end{abstract}

\section{Introduction}
\textbf{Markovian evolving graphs and Flooding.} Graphs that evolve over time are currently a very hot topic in computer science. They arise from several areas such as mobile networks, networks of users exchanging e-mail or instant messages, citation networks and hyperlinks networks, peer-to-peer networks, social networks (who-trust-whom, who-talks-to-whom, etc.), and many other more~\cite{AKL08,BOOK05,CMPS07,OS07,HHL88,P87,Sh01}.

\emph{Markovian evolving graphs} are a natural and very general class of models for evolving graphs introduced in \cite{AKL08}. In these models, the set of nodes is fixed and the edge set at time $t$ stochastically depends on the edge set at time $t-1$: so, we have an infinite sequence of graphs that is a Markov chain. It is important to observe that, on one hand, these models make the underlying mechanism of how the graph evolves explicit; on the other hand, they are very general since, by a suitable choice of the \emph{matrix transition probability} yielding the graph Markovian process, it is possible to model several important network scenarios such as \emph{faulty-networks} and \emph{geometric-mobile} networks (such scenarios will be described later).

In \cite{AKL08}, the hitting time and cover time of random walks in some specific cases of Markovian evolving graphs have been analytically studied. We instead investigate the speed of \emph{information spreading} on general Markovian evolving graphs. Reaching all nodes from a given source/initiator node is typically required to disseminate or retrieve information: this task is performed via suitable protocols that aim to achieve low delay and message overhead. However, when the network topology is highly dynamic and unknown, (e.g. unstructured peer-to-peer networks, faulty/mobile networks, etc), it is very hard to design efficient protocols   and, as a result, the \emph{flooding mechanism} is often adopted \cite{CL05,GS05,G02,LCS02}. In the flooding mechanism, any \emph{informed} node (i.e. any node that has the source message) always sends the source message to all its neighbors. So, the source is informed since the beginning and, clearly, any other node gets informed at time step $t$ iff any of its neighbors (w.r.t. the edge set at time $t$) is informed at time step $t-1$. \\
The completion time of the flooding mechanism (shortly \emph{flooding time}) is the first time step in
which all nodes of the network are informed.

\noindent It is important to observe that flooding time of a  dynamic network  may largely differ from its diameter: for instance, it is easy to construct  an $n$-node mobile  network over a finite square that has, at every time, diameter $D =3$ while its flooding time is $\Theta(n)$. In general, any diameter bound for a given dynamic network implies nothing about its flooding time but the fact that the latter is finite. Flooding time in fact represents the ``natural'' lower bound for broadcast protocols in dynamic networks. For this reason, flooding is often used in order to evaluate the relative efficiency of alternative protocols, especially in networks with unknown dynamic topology \cite{CL05,GS05,OS07}.

\medskip \noindent \textbf{Our results.}
We study flooding time in \emph{stationary} Markovian evolving graphs, i.e., when the \emph{initial} graph is random with a \emph{stationary distribution} of the underlying Markov chain \cite{AF99}. In network mobility simulation, this corresponds to the important concept of \emph{perfect simulation} (see \cite{LV05,CNB04}).

We prove an upper bound on flooding time in \emph{any} stationary Markovian evolving graph. This upper bound is expressed in terms of the parameterized node-expansion properties satisfied by the stationary graphs. As far as we know, this is the first analytical result on the speed of information spreading in so general dynamic models.

\noindent We then show the tightness (so the ``goodness'') of this bound in two relevant concrete network scenarios: \emph{geometric Markovian evolving graphs} (in short, \emph{geometric-MEG}) and \emph{edge-Markovian evolving graphs} (in short, \emph{edge-MEG}).

\smallskip \noindent \emph{Geometric Markovian evolving graphs.}
We consider a model of evolving graphs that is based on \emph{node mobility}. It is the discrete version of the well-known \emph{random-walk} model \cite{G87,CBD02,DMP08,JMR09}. In this model, denoted here as \emph{geometric-MEG}, nodes (i.e. radio stations) move over a region of the plane (typically a square region) and each node performs, independently from the others, a sort of Brownian motion. At any time there is an edge (i.e. a bidirectional connection link) between two nodes if they are at distance at most $R$ (typically $R$ represents the transmission range). We make time discrete and consider a square grid of arbitrary resolution as a node support-space (see Section~\ref{sec::geom} for details). This model can also be viewed as the \emph{walkers} model \cite{DMP08} on the square grid.

\noindent It is important to observe that geometric-MEG yield stochastic dependency among the dynamic edges, i.e., the probability of an edge depends on the existence of other edges.

\noindent The impact of mobility in information spreading has been the subject of several papers over the last years.
However, only few analytical results are currently available. In \cite{GT02}, some bounds on the network capacity (i.e. the number of received packets) has been analyzed on a mobility model that is not explicitly defined. In \cite{KY08}, the authors analyze the broadcast time over a restricted mobility model. In this restricted model, at every time step, the position of each  node is selected independently at random inside a disk that is fixed at the starting time. Observe that, in this restriction, there is no stochastic dependence between two consecutive node positions: the model is significantly far from the random walk model. Then, the same work provides some experimental results for the random walk model. Finally, in \cite{JMR09}, the speed of data communication between two nodes is studied over a class of \emph{Random-Direction} models yielding uniform stationary node distributions (including the random walk model \emph{with reflection}). They provide an upper bound on this speed that can be interpreted as a   \emph{lower} bound on flooding (routing) time when the mobile  network is very sparse and disconnected (so, differently from our result, they consider geometric-MEG \emph{under} the connectivity threshold). Their adopted technique based on \emph{Laplacian transform of independent journeys} strongly departs from ours and it cannot be extended to provide any upper bound on flooding time. Further related analytical results that have been obtained after the conference version of our work are discussed in Section~\ref{sec::concl}.

\noindent We first prove that stationary geometric MEG, yielding \emph{connected} graphs, satisfy certain parameterized node-expansion properties. We then apply our general result and achieve an upper bound on flooding time. The obtained bound is shown to be tight whenever flooding time is $\Omega(\log\log n)$. Informally speaking, this happens whenever \emph{(i)} the transmission radius is \emph{not} ``almost'' equal to the diameter of the square region and \emph{(ii)} the maximal \emph{node speed} is less than the \emph{message-transmission} speed. Both assumptions are satisfied by most of real mobile networks. In general, our upper bound is thus at most an $O(\log\log n)$ additive factor larger than the optimum.

\smallskip \noindent \emph{Further mobility models.} The node-expansion properties of geometric-MEG are mainly due to the fact that the stationary distribution of node positions is \emph{almost uniform}. In this paper, we provide formal results and proofs only for flooding in geometric-MEG. However, our \emph{expansion} technique can be applied to \emph{any} mobility model yielding a uniform or almost uniform stationary distribution of node positions. Several variants of the \emph{random waypoint} model, one of the most commonly used mobility models \cite{JN96,CBD02,LV05}, enjoy this uniformity property. Among the others, we mention the random-direction model \emph{with reflection} (also called the \emph{billiard } model) \cite{BL03,LV05,NTLL05}, the random waypoint \emph{on a torus} \cite{G87,H97,LV05,NTLL05} and the random waypoint \emph{on a sphere} \cite{LV05}. Furthermore, the uniformity property is also satisfied by the \emph{walkers} model on a toroidal grid \cite{DPSW08}.

To the best of our knowledge, our results are the first \emph{analytical} bounds on flooding time for natural and concrete models of \emph{mobile} networks.

We finally remark that our flooding analysis does not take care of the \emph{interference} problem in message transmissions: this is typically managed at the MAC layer of a wireless network architecture \cite{BL03,CMPS07}. The impact of message interferences in geometric-MEG is a further interesting issue which is out of the scope of our work focussing instead on \emph{dynamic-topological} properties of MEG.

\smallskip \noindent \emph{Edge-Markovian evolving graphs.}
In several network scenarios, there is a strong dependence between the existence (or the absence) of a link between two nodes at a given time step and the existence (or the absence) of the same link at the previous time step. Important examples of this behavior arise in faulty communication networks, peer-to-peer networks\footnote{Notice that, in some of these settings, there is an underlying physical network that supports the abstraction of point-to-point communication.}, and social networks.

We thus consider \emph{edge-MEG}, special Markovian evolving graphs, recently studied in \cite{CMMPS08,BCF09}, which are a time-discrete version of the \emph{reciprocity graph model} introduced in the context of evolving social networks \cite{W80}. At every time step, every edge changes its state (existing or not) according to a two-state Markovian process with probabilities $p(n)$ and $q(n)$ where $n$ is the number of nodes. If an edge exists at time $t$ then at time $t+1$ it dies with probability $q(n)$ (i.e. \emph{death-rate}). If instead the edge does not exist at time $t$, then it will come into existence at time $t+1$ with probability $p(n)$ (i.e. \emph{birth-rate}). For brevity's sake, functions $p(n)$ and $q(n)$ will be simply denoted as $p$ and $q$, respectively\footnote{Hence, any inequality $p \leqslant (\geqslant) b(n)$ means that $p(n)$ is eventually not larger (not smaller) than $b(n)$. The same holds for $q = q(n)$.}. Observe that setting $q = 1-p$ yields (time-independent) \emph{dynamic random graphs}   studied in \cite{CMPS07} to model dynamic radio networks and in \cite{BDW08} to model  epidemic biological  processes; here   links, at every time, are chosen independently at random. So, edge-MEG are (in turn) a wider and more realistic class of dynamic random graphs. Observe that when $0 < p,q <1$, the stationary distribution is unique.

Similarly to the case of geometric-MEG, we first prove that stationary edge-MEG, yielding \emph{connected} graphs, satisfy certain parameterized node-expansion properties. Thanks to these properties, we can apply our general result and achieve an upper bound on flooding time. The obtained bound is shown to be tight whenever flooding time is $\Omega(\log\log n)$: this includes, for instance, the relevant case where the expected node-degree is $O(\polylog n)$. In general, our upper bound for edge-MEG is thus at most an $O(\log\log n)$ additive factor larger than the optimum.

In \cite{CMMPS08}, the \emph{maximal} flooding time has been studied  in edge-MEG with respect to \emph{any} initial probability distribution. In that paper, in fact, almost tight bounds for the \emph{worst-case} flooding time have been derived. However, those results do not say whether flooding can be (significantly) faster in \emph{stationary} edge-MEG. Interestingly enough, our stationary bound implies that, whenever the birth-rate $p$ is $O(1/n^{1+\epsilon})$ and the death-rate $q$ is $O(np/\log n)$, there is an \emph{exponential} gap between the stationary case and the worst-case. An exponential gap also holds whenever $p = O(\log n/n)$ and $q = O(p \sqrt n)$ (for instance, set $q = \polylog n /n$).

\medskip \noindent \textbf{Organization of the paper.}
In Section~\ref{sec::gen}, we prove our upper bound for flooding time in general Markovian evolving graphs. The results for geometric-MEG and edge-MEG are described in Sections~\ref{sec::geom} and~\ref{sec::edge}, respectively. Finally, further related analytical results (obtained after the conference version of our work) and some open questions are discussed in Section~\ref{sec::concl}.

\section{Markovian evolving graphs: the general theorem}\label{sec::gen}
Through this paper, the set  $[n]=\{ 1, \ldots , n\}$ will represent the set of $n$ nodes. Let $G=([n],E)$ be a graph and $I \subseteq [n]$ be a subset of nodes. We denote by $N(I)$ the out-neighborhood of $I$, i.e.
$$
N(I) = \{ v \in  [n] \setminus I \,:\, \{ u,v \} \in E,\, \mbox{ for some } u \in I \}
$$
Given a source node $s \in [n]$, the \emph{flooding} process can be represented by the sequence $\{I_t \subseteq [n] \,:\, t \in \mathbb{N}\}$ where $I_t$  is the subset of informed nodes defined recursively as follows
\[
\left\{
\begin{array}{lcl}
I_0 & = & \{s\}  \\
I_{t+1} & = & I_t \cup N(I_t)
\end{array}
\right.
\]
Notice that the subset $N(I_t)$ refers to the graph at time step $t$. Let $T(s)$ be the first time step such that all nodes are informed. The \emph{flooding time} is the maximum $T(s)$ over all possible choices of source $s$.

\begin{definition}[Markovian evolving  graph]\label{def::Markovevolv}
Let $\mathbf{G}$ be a family of graphs with the same node set $[n]$. A \emph{Markovian evolving graph} $\mathcal{M} =  \{ G_t \,:\, t \in \mathbb{N} \} $ is a Markov chain with state space $\mathbf{G}$.

\noindent A \emph{stationary} Markovian evolving  graph is a Markovian evolving graph $\mathcal{M} = \{ G_t \,:\, t \in \mathbb{N} \}$ such that $G_0$ is random with a stationary distribution of $\mathcal{M}$.
\end{definition}

\noindent The following definition concerns a sort of parameterized node-expansion. This is a key-ingredient, in our analysis of flooding in Markovian evolving  graphs, to cope with the difficulties due to the stochastic dependence.

\begin{definition}[Expander]
A graph $G=([n],E)$ is a \emph{$(h,k)$-expander} if, for every set of nodes $I \subseteq [n]$ with $|I| \leqslant h$, it holds that $|N(I)| \geqslant k |I|$.
\end{definition}

\noindent The above definition naturally extends to random variables and their probability distributions.

\begin{definition}[Expander II]
Let $X$ be a random variable with values in a family of graphs with the same node set $[n]$. Then $X$ is a \emph{$(h,k)$-expander} with probability $p$ if
$$
\Prob{X \mbox{ is a } (h,k)\mbox{-expander}} \geqslant p
$$
In this case, we also say that the probability distribution of $X$ \emph{yields} an $(h,k)$-expander with probability $p$.
\end{definition}

\noindent We are now able to provide our main result for general stationary Markovian evolving graphs. We first show a lemma that connects the parameterized expansion of a (deterministic) evolving graph with its flooding time, then we use it to prove our theorem on the flooding time of stationary Markovian evolving graphs. In what follows, all    logarithms are in  base $e$.

\begin{lemma}[Flooding and Expansion: Deterministic case]\label{lemma:detbig}
Let $\mathcal{G} = \{ G_t \,:\, t \in \mathbb{N} \}$ be an evolving graph (i.e. a sequence of graphs with the same node set $[n]$). Suppose an increasing sequence $1 = h_0 \leqslant h_1 < \dots < h_s = n/2$ and a non-increasing sequence $k_1 \geqslant  \dots \geqslant k_s$ of positive real numbers exist such that, for every $t \in \mathbb{N}$, graph $G_t$ is a $(h_i,k_i)\mbox{-expander}$ for every $i=1, \dots, s$. Then the flooding time of $\mathcal{G}$ is
$$
\mathcal{O}\left( \sum_{i=1}^s \frac{\log (h_i/h_{i-1})}{\log (1 + k_i)} \right)
$$
\end{lemma}
\proof Let $m_t$ be the number of informed nodes at time step $t$, at the beginning $m_0 = 1$. For $i = 1, \dots, s$ let $T_i$ be the first time step such that the number of informed nodes is larger than $h_i$,
$$
T_i = \min \{ t \in \mathbb{N} \;:\; m_t > h_i \}
$$
If $t$ is a time step such that $h_{i-1} < m_t \leqslant h_i$ for some $i = 1, \dots, s$, then, since the graph $G_t$ is a $(h_i, k_i)$-expander, it holds that the number of informed nodes at the next time step is
$$
m_{t+1} \geqslant (1 + k_i) m_t
$$
Hence, after $t'$ of such time steps it holds that
$$
m_{t+t'} \geqslant (1 + k_i)^{t'} m_t \geqslant (1 + k_i)^{t'} h_{i-1}
$$
So the number of time steps between $T_{i-1}$ and $T_i$ is at most
$$
T_i - T_{i-1} \leqslant \left\lceil \frac{\log(h_i/h_{i-1})}{\log(1 + k_i)} \right\rceil
$$
If $\frac{\log(h_i/h_{i-1})}{\log(1 + k_i)} = \Omega(1)$ then $T_i - T_{i-1} = \mathcal{O}\left( \frac{\log(h_i/h_{i-1})}{\log(1 + k_i)} \right)$. If $\frac{\log(h_i/h_{i-1})}{\log(1 + k_i)} = o(1)$ then $(1+k_i)h_{i-1} \gg h_i$, that is in just one time step the number of informed nodes \emph{jumps} from $m_t \leqslant h_i$ to $m_{t+1}$ much larger than $h_i$. Let $j$ be the index such that $h_j < (1+k_i)h_{i-1} \leqslant h_{j+1}$ (if none of such index exists it means $m_{t+1} > n/2$, so $T_s - T_{i-1} = 1$), then it holds that
$$
1 \leqslant \frac{\log(h_{j+1}/h_{i-1})}{\log(1 + k_i)} = \sum_{\ell = i}^{j+1} \frac{\log(h_\ell/h_{\ell-1})}{\log(1 + k_i)} \leqslant \sum_{\ell = i}^{j+1} \frac{\log(h_\ell/h_{\ell-1})}{\log(1 + k_\ell)}
$$
In the last inequality we used that $k_\ell \leqslant k_i$ for $\ell \geqslant i$. Hence we can bound 
$$
T_j - T_{i-1} \leqslant \left\lceil \frac{\log(h_{j+1}/h_{i-1})}{\log(1 + k_i)} \right\rceil
= \mathcal{O}\left( \sum_{\ell = i}^{j+1} \frac{\log(h_\ell/h_{\ell-1})}{\log(1 + k_\ell)} \right)
$$
Finally, by summing up the contributions of all the considered time intervals we have
$$
T_s = \mathcal{O}\left( \sum_{i=1}^s \frac{\log (h_i/h_{i-1})}{\log (1 + k_i)} \right)
$$

\smallskip\noindent Once we have at least $n/2$ informed nodes, then a symmetric argument holds. Indeed, consider the number $\bar{m}_t$ of \emph{non-informed} nodes at time step $t$. Observe that the neighbors, in graph $G_t$, of such nodes were not informed at the previous time step $t-1$ (otherwise at time step $t$ they would have informed their neighbors in graph $G_t$). Let $i \in \{0, 1, \dots, s\}$ be the index such that $h_{i-1} < \bar{m}_t \leqslant h_i$, since $G_t$ is a $(h_i,k_i)$-expander the number of such neighbors is at least $k_i \bar{m}_t$, so the number of non-informed nodes at the previous time step $t-1$ were at least
$$
\bar{m}_{t-1} \geqslant (1+k_i) \bar{m}_t
$$
In other words, the number of non-informed nodes follows the same growth rate, when there are at least $n/2$ informed nodes and we look at the system going backward in time, of the number of informed nodes, when there are less than $n/2$ informed nodes and the time moves forward. Hence, to go from $n/2$ informed nodes to $n$ informed nodes it takes further $\mathcal{O}\left( \sum_{i=1}^s \frac{\log (h_i/h_{i-1})}{\log (1 + k_i)} \right)$ time steps.
\qed

As usual, we say that an event $\mathcal(E)(n)$ holds \emph{with high probability} (for short \emph{w.h.p.}) if 
 $\Prob(\mathcal(E)(n)) \geqslant 1- 1/n$.
\begin{theorem}\label{theorem:big}
Let $\mathcal{M} = \{ G_t \,:\, t \in \mathbb{N} \}$ be a stationary Markovian evolving graph. Assume an increasing sequence $1 = h_0 \leqslant h_1 < \dots < h_s = n/2$ and a non-increasing sequence $k_1 \geqslant \dots \geqslant k_s$ (for any $s \leq n/2$) of positive real numbers exist such that, with probability $1 - 1/n^2$, for every $i=1, \dots, s$ the stationary distribution of $\mathcal{M}$ yields an $(h_i,k_i)\mbox{-expander}$. Then the flooding time of $\mathcal{M}$ is w.h.p.
$$
\mathcal{O}\left( \sum_{i=1}^s \frac{\log (h_i/h_{i-1})}{\log (1 + k_i)} \right)
$$
\end{theorem}
\proof
For $t = 0, 1, \dots$ define the event
$$
\mathcal{E}_t = \left\{ \, G_t \mbox{ is a } (h_i,k_i)\mbox{-expander for every } i = 1, \dots, s \, \right\}
$$
By stationarity hypothesis we have that $\Prob{\mathcal{E}_t} \geqslant 1 - 1/n^2$ for every $t$. Now consider the event
$$
\mathcal{F} = \left\{ \, \mbox{The flooding time of } \mathcal{M} \mbox{ is } \mathcal{O}\left( \sum_{i=1}^s \frac{\log (h_i/h_{i-1})}{\log (1 + k_i)} \right) \, \right\}
$$
and observe that from Lemma~\ref{lemma:detbig} it follows that $\bigcap_{t=0}^n \mathcal{E}_t \subseteq \mathcal{F}$, hence $\overline{\mathcal{F}} \subseteq \bigcup_{t=0}^n \overline{\mathcal{E}_t}$ and we have that
$$
\Prob{\overline{\mathcal{F}}} \leqslant \Prob{\bigcup_{t=0}^n \overline{\mathcal{E}_t}} \leqslant n \Prob{\overline{\mathcal{E}_0}} \leqslant \frac{1}{n}
$$
\qed

\smallskip\noindent
An easy consequence of Theorem~\ref{theorem:big} is the  following

\begin{cor}\label{cor:big}
Let $\mathcal{M} = \{ G_t \,:\, t \in \mathbb{N} \}$ be a stationary Markovian evolving graph.  Assume
a non-increasing sequence $k_1 \geqslant \dots \geqslant k_{n/2}$ of positive real numbers exists such
that, with probability $1 - \frac{1}{n^2}$, for every $i=1, \dots, n/2$  the stationary distribution of $\mathcal{M}$ yields an
$(i,k_i)\mbox{-expander}$. Then the flooding time of $\mathcal{M}$ is
w.h.p.
$$
\mathcal{O}\left( \sum_{i=1}^{n/2} \frac{1}{i \, \log (1 + k_i)} \right)
$$
\end{cor}

\section{Geometric Markovian evolving graphs} \label{sec::geom}

We introduce a model of dynamic graphs that is a discrete version of the \emph{random walk mobility} model for radio networks \cite{CBD02}. In the latter model, nodes (i.e. radio stations) move on a bounded region of the plane (typically a square region) and each node performs, independently from the others, a sort of Brownian motion. At any time there is an edge (i.e. a bidirectional connection link) between two nodes if they are at distance at most $R$ (typically $R$ represents the transmission range). In our model we discretize time and space. We choose to keep constant the density (i.e. the ratio between the number of nodes and the area) as the number $n$ of nodes grows. The node region is a square of side $\sqrt{n}$ and the density equals to 1. This choice is only for the sake of
simplicity and all the results can be scaled to any density $\delta(n)$ (see Observation \ref{obs::delta}). 
The nodes can assume positions whose coordinates are integer multiple of a sufficiently small resolution coefficient $\epsilon >0$; in the sequel, we always assume that $\epsilon \leqslant  1$  and $\epsilon < R$.
Formally, nodes move on the following set of points
$$
L_{n,\epsilon} = \{ (i\epsilon, j\epsilon)\;|\; i,j\in \mathbb{N} \wedge i, j \leqslant \frac{\sqrt{n}}{\epsilon}\}
$$
At any time step, a node can move to one of the positions of $L_{n,\epsilon}$ within distance $r$ from the previous position. The positive real number $r$ is a fixed parameter that we call \emph{move radius}. It can be interpreted as the maximum velocity of a node\footnote{Indeed, a node can run through a distance of at most $r$ in a unit of time.}. Formally, we introduce the \emph{move graph} $M_{n, r, \epsilon} = (L_{n,\epsilon}, E_{n,r,\epsilon})$, where
$$
E_{n,r, \epsilon} = \{ (\mathbf{x}, \mathbf{y}) \;|\; \mathbf{x}, \mathbf{y}\in L_{n,\epsilon}\;\;
d(\mathbf{x}, \mathbf{y}) \leqslant r \}
$$
and $d(\cdot, \cdot)$ is the Euclidean distance. A node in position $\mathbf{x}$, in one time step, can move in any position in $\Gamma(\mathbf{x})$, where $\Gamma(\mathbf{x}) = \{ \mathbf{y}\; |\; (\mathbf{x}, \mathbf{y})\in E_{n,r,\epsilon}\}$. The nodes are identified by the first $n$ positive integers $[n]$. The time-evolution of the movement of a single node $i$ is represented by a Markov chain $\{P_{i,t}\; ;\; t\in\mathbb{N}\}$ where $P_{i,t}$ are random variables whose state-space is $L_{n,\epsilon}$ and
$$
\Prob{P_{i,t+1} = \mathbf{x} } \, = \,
\left\{
\begin{array}{ll}
\frac{1}{|\Gamma(P_{i, t})|} & \mbox{if $\mathbf{x} \in \Gamma(P_{i, t})$}\\
0 & \mbox{otherwise}
\end{array}
\right.
$$
In other words, $P_{i,t}$ is the position of node $i$ at time $t$. Thus, the time-evolution of the movements of all the nodes is represented by a Markov chain $\mathcal{P}(n, r, \epsilon) = \{P_t \: :\; t\in \mathbb{N} \}$ whose state-space is $L_{n,\epsilon}\times L_{n,\epsilon}\times \cdots \times L_{n,\epsilon}$ ($n$ times) and
$$
P_t  = (P_{1,t}, P_{2,t},\ldots, P_{n, t})
$$
Let us fix a \emph{transmission radius} $R > 0$. A \emph{geometric-MEG} is a sequence of random
variables $\mathcal{G}(n, r, R, \epsilon) = \{ G_t \;:\; t \in \mathbb{N}\}$ such that $G_t = ([n],
E_t)$ with
$$
E_t = \{ (i, j) \;|\; d(P_{i, t}, P_{j, t}) \leqslant R\}
$$
From a formal point of view, geometric-MEGs are not Markovian evolving graphs according to Definition~\ref{def:genmg}. In order to include them, we  need a slight generalization of the definition

\begin{definition}[Markovian Evolving  Graph II]\label{def:genmg}
Let $\mathbf{G}$ be a family of graphs with the same node set $[n]$. A \emph{Markovian evolving  graph} $\mathcal{G} =  \{ G_t \,:\, t \in \mathbb{N} \} $ is a sequence of random variables with state space $\mathbf{G}$ and such that there exist both  a Markov chain $\mathcal{X} =  \{ X_t \,:\, t \in \mathbb{N} \}$ and a function $f$ so that $G_t = f(X_t )$. \\
A \emph{stationary} Markovian evolving  graph is a Markovian evolving graph $\mathcal{G} = \{ G_t \,:\, t \in \mathbb{N} \}$ such that $G_0$ is random with a stationary distribution of $\mathcal{X}$ translated by $f$.
\end{definition}

\noindent Theorem \ref{theorem:big} easily extends to the above generalized definition by straightforward arguments.

\noindent As for the stationary case, standard results of Markov chain theory (see~\cite{AF99}) easily imply that the (unique) stationary distribution $\pi_i$ of Markov chain $\{P_{i,t}\; ;\; t\in\mathbb{N}\}$ is
$$
\pi_i(\mathbf{x}) \, = \, \frac{|\Gamma(\mathbf{x})|}{\sum_{\mathbf{y}\in L_{n,\epsilon}} |\Gamma(\mathbf{y})|}
$$
Notice that $\pi_i$ is almost uniform since, for any two  positions $\mathbf{x}$ and $\mathbf{y}$, the values $\pi_i(\mathbf{x})$ and $\pi_i(\mathbf{y})$ can differ by at most a constant factor. Moreover, the stationary distribution of $\mathcal{P}(n, r, \epsilon)$ is the product of the independent distributions $\pi_i$ for all $i\in [n]$. We say that a geometric-MEG $\mathcal{G}(n, r, R, \epsilon) = \{ G_t \;:\; t \in \mathbb{N}\}$ is a \emph{stationary} geometric-MEG if the underlying $P_0$ is random with the stationary distribution of the Markov chain $\mathcal{P}(n, r, \epsilon) = \{P_t \: :\; t\in \mathbb{N} \}$. Notice that if $\mathcal{G}(n, r, R, \epsilon) = \{ G_t \;:\; t \in \mathbb{N}\}$ is a stationary geometric-MEG then all random variables $G_t$ are random with the same probability distribution that we call \emph{stationary distribution} of $\mathcal{G}(n, r, R, \epsilon)$.

\noindent Stationary geometric-MEG enjoy of the following expansion properties.

\begin{theorem}\label{thm:geomexpan}
If  $r\geqslant 0$ and  $c\sqrt{\log n} \leqslant R \leqslant \sqrt{n}$ for a sufficiently large constant $c$, then constants $\alpha, \beta > 0$ exist such that, with probability $1 - \frac{1}{n^2}$, the stationary distribution of $\mathcal{G}(n, r, R, \epsilon)$ yields:
\begin{itemize}
\item A $(h, \alpha\frac{R^2}{h})$-expander for $1 \leqslant h \leqslant \alpha R^2$;
\item A $(h, \beta\frac{R}{\sqrt{h}})$-expander for $\alpha R^2 \leqslant h \leqslant n/2$.
\end{itemize}
\end{theorem}

\begin{obs}\label{obs::delta}
For general node density $\delta(n)$, Theorem \ref{thm:geomexpan} holds under the scaled assumption $R \geqslant c \sqrt{\log n / \delta(n)}$.
\end{obs}
\proof
Let $m = \lceil\sqrt{5n}/R\rceil$. Consider the partition of the square $\sqrt{n}\times\sqrt{n}$ into $m\times m$ congruent sub-squares, called cells. Every cell can be identified by the pair of indices $(i, j)$, for $1\leqslant i, j\leqslant m$, such that $i$ is the index of row and $j$ is the index of column of the cell. Let $c_{i,j}$ be the subset of the points of $L_{n,\epsilon}$ that fall into the cell $(i, j)$. Notice that the side length $\ell$ of a cell satisfies
\[
R/(\sqrt{5} + 1) \leqslant \ell\leqslant R/\sqrt{5}
\]
Thus, any point of a cell is at distance less than $R$ from any point of a side-by-side adjacent cell.

\noindent Through the following, we assume that the positions of the nodes are random with the stationary distribution of the Markov chain $\mathcal{P}(n, r, \epsilon)$. Moreover, we say that a node belongs to a cell whenever its position belongs to the cell. \noindent Let $N_{i,j}$ be the random variable counting the number of nodes in cell $c_{i,j}$. Now, we prove a simple but crucial claim.

\begin{quote}
\begin{claim}\label{claim:density}
If $\epsilon \leqslant 1$ and $R \geqslant c\sqrt{\log n}$ for a sufficiently large constant $c$, then a constant $\lambda \geqslant 1$ exists such that, with probability $1 - \frac{1}{n^2}$, it holds that, for every $1\leqslant i, j\leqslant m$,
$$
\frac{R^2}{\lambda} \, \leqslant \, N_{i,j} \, \leqslant \, \lambda R^2
$$
\end{claim}
\proof Firstly, consider a fixed cell $(i,j)$. For every $u \in [n]$, let $X_u$ be the $\{0,1\}$
random variable that is 1 iff node $u$ is in the cell $c_{i,j}$. Clearly, these are independent random
variables and it holds that $N_{i,j} = \sum_{u\in [n]} X_u$. As for the probability distribution of
$X_u$, we have that
\[
\Prob{X_u = 1} \, = \, \sum_{\mathbf{x}\in c_{i,j}} \pi_u(\mathbf{x})
\]
Since
\[
\pi_u(\mathbf{x}) = \frac{|\Gamma(\mathbf{x})|}{\sum_{y\in L_{n,\epsilon}}|\Gamma(\mathbf{y})|}
\]
it is easy to see that, if $\epsilon$ is sufficiently small (say $\epsilon\leqslant 1$) then there is a constant $\gamma \geqslant 1$ such that, for every $\mathbf{x} \in L_{n,\epsilon}$, it holds that
\[
\frac{1}{\gamma |L_{n,\epsilon}|} \,\leqslant\, \pi_u(\mathbf{x}) \, \leqslant \,
\frac{\gamma}{|L_{n,\epsilon}|}
\]
This implies that
\[
\frac{|c_{i,j}|}{\gamma |L_{n,\epsilon}|} \, \leqslant \, \Prob{X_u = 1}\, \leqslant \, \frac{\gamma |c_{ij}|}{|L_{n,\epsilon}|}
\]
By taking into account the side length of the cells, it is easy to verify that 
\[
\frac{R^2}{10n}\, \leqslant \, \frac{|c_{i,j}|}{|L_{n,\epsilon}|} \, \leqslant\, \frac{2 R^2}{5n}
\]
It follows that
\[
\frac{R^2}{10\gamma n}   \leqslant \Prob{X_u = 1} \leqslant \frac{2\gamma R^2}{5n}
\ \mbox{ and } \
\frac{R^2}{10\gamma} \leqslant \Expec{N_{i,j}} \leqslant \frac{2\gamma R^2}{5}
\]
Now, in virtue of the Chernoff's bound \cite{ALICE}, if $R \geqslant c\sqrt{\log n}$, for a sufficiently large
constant $c$, then a constant $\lambda \geqslant 1$ exists such that
\[
\frac{R^2}{\lambda} \, \leqslant\, N_{i,j} \, \leqslant \,  \lambda R^2
\]
with probability at least $1 - \frac{1}{n^3}$. Since the number of cells is less than $n$, a simple
application of the union bound proves the thesis of the claim.
\qed
\end{quote}

\noindent Let $\mathcal{B}$ be the event that occurs when, for every $1\leqslant i, j\leqslant m$,
$$
\frac{R^2}{\lambda} \, \leqslant \, N_{i,j} \, \leqslant\, \lambda R^2
$$
where $\lambda$ is the constant of Claim~\ref{claim:density}. We now prove event $\mathcal{B}$ implies
the expansion properties stated in the thesis of the theorem.

\begin{quote}
\begin{claim}\label{claim:exp1}
If event $\mathcal{B}$ holds then the graph induced by $R$ and by the positions of the nodes is a $(h,
\alpha\frac{R^2}{h})$-expander for $1 \leqslant h \leqslant \alpha R^2$, where $\alpha =
1/(2\lambda)$.
\end{claim}
\proof Let $I \subseteq [n]$ be such that $|I| \leqslant \alpha R^2$. Consider a node  $u$ in $I$ and
let $c_{i,j}$ be the cell that contains $u$. Since $\mathcal{B}$ holds, $N_{i,j} \geqslant
\frac{R^2}{\lambda}$. All the nodes in $c_{i,j}$ are adjacent to $u$. Thus, there are at least
$N_{i,j} - |I|$ nodes that are adjacent to $u$ and that are not in $I$. It follows that
\begin{small} $$
|N(I)|  \, \geqslant \,  N_{i,j} - |I| \, \geqslant \, \frac{R^2}{\lambda} - \alpha R^2 \, \geqslant \,
\frac{R^2}{2\lambda} \, = \, \alpha R^2
$$ \end{small}
In other terms, $|N(I)| \geqslant \alpha\frac{R^2}{|I|} |I|$.
\qed

\begin{claim}\label{claim:exp2}
If event $\mathcal{B}$ holds then the graph induced by $R \leqslant \sqrt{n}$ and by the positions of
the nodes is a $(h, \beta\frac{R}{\sqrt{h}})$-expander for $\alpha R^2 \leqslant h \leqslant n/2$,
where $\beta = \frac{1}{8\lambda^2}$.
\end{claim}
\proof Let $I \subseteq [n]$ be any subset of nodes with $|I| \leqslant n/2$. We say that a cell is
\emph{black} if it contains at least one node in $I$. Let $B$ be the random variable that is the set of \emph{black} cells. Let $J$ be the random variable defined as follows
$$
J = \{ u\in [n]\;|\; u\not\in I \;\wedge\; \exists c\in B\;:\; \mbox{node $u$ belongs to $c$}\}
$$
Now, two cases are possible: either $|J| \geqslant \beta R \sqrt{|I|}$ or not. Firstly, suppose that
$|J| \geqslant \beta R \sqrt{|I|}$. Since every node in $J$ is in a \emph{black} cell, it holds that
$J \subseteq N(I)$, and thus
\begin{small} $$
|N(I)| \, \geqslant \, |J| \, \geqslant \, \beta R \sqrt{|I|}
$$ \end{small}
In other terms, $N(I) \geqslant \beta\frac{R}{\sqrt{|I|}}|I|$ and the expansion property is proved.

\noindent Consider now the case $|J| < \beta R \sqrt{|I|}$. We say that a row (column) of cells is
\emph{black} if all the cells of the row (column) are \emph{black}. Similarly, we say that a row
(column) is \emph{white} if all the cells of the row (column) are \emph{white}. A row (column) that is
neither \emph{black} nor \emph{white} is said to be \emph{gray}. Notice that a \emph{gray} row
(column) contains at least two adjacent cells such that one is \emph{non-black} and the other is
\emph{black}. Let $B_r$ and $B_c$ be, respectively, the number of \emph{black} rows and the number of
\emph{black} columns. Three cases may arise.

\smallskip \noindent
[$B_r \geqslant 1$]: Observe that in this case all the columns are either \emph{black} or \emph{gray}.
Let $Y$ be the number of \emph{gray} columns.  It holds that
$$
Y \, \geqslant \,  m - B_c \, \geqslant \, m - \frac{|B|}{m}
$$
Since event $\mathcal{B}$ holds, the number of nodes in non-\emph{black} cells is bounded by $\lambda
R^2 (m^2 - |B|)$ and thus
$$
\lambda R^2 (m^2 - |B|) \, \geqslant \, n - |I| - |J| \, \geqslant \, n - |I| - \beta R \sqrt{|I|}
$$
It follows that
$$
|B| \, \leqslant \, m^2 - \frac{n - |I| - \beta R \sqrt{|I|}}{\lambda R^2}
$$
By combining this bound with the previous bound on $Y$ we obtain
$$
Y \, \geqslant \, \frac{n - |I| - \beta R \sqrt{|I|}}{\lambda R^2m} \, \geqslant \, \frac{n - |I| - \beta
R \sqrt{|I|}}{\lambda 2\sqrt{5n} R}
$$
where the last inequality follows from $m = \lceil\sqrt{5n}/R\rceil$ and $R \leqslant \sqrt{n}$.

\noindent Observe that every \emph{gray} column contains at least one \emph{non-black} cell that is adjacent
to a \emph{black} cell. So, all the nodes belonging to those \emph{non-black} cells are included in
$N(I)$. Since event $\mathcal{B}$ holds, it follows that
$$
|N(I)| \, \geqslant \, Y \frac{R^2}{\lambda} \, \geqslant \, R\left(\frac{n - |I| - \beta R
\sqrt{|I|}}{\lambda^2 2\sqrt{5n}}\right)
$$
Now, recalling that $\beta = \frac{1}{8\lambda^2}$, $|I| \leqslant n/2$, and $R \leqslant \sqrt{n}$, it is easy to verify that
$$
\frac{n - |I| - \beta R \sqrt{|I|}}{\lambda^2 2\sqrt{5n}} \geqslant \beta\sqrt{|I|}
$$
It follows that $|N(I)| \geqslant \beta R \sqrt{|I|}$ and the expansion property holds.

\smallskip \noindent
[\mbox{\rm $B_c \geqslant 1$ (and $B_r = 0$)}]: This case is symmetric to the previous one.

\smallskip \noindent
[\mbox{\rm $B_r = 0$ and $B_c = 0$}]: In this case, all the rows and columns are either \emph{gray} or
\emph{white}. Let $Y_r$ and $Y_c$ be the number of \emph{gray} rows and the number of \emph{gray}
columns, respectively. Since there are neither \emph{black} rows nor \emph{black} columns, it must be
the case that every \emph{black} cell belongs to both a \emph{gray} row and a \emph{gray} column. As a
consequence it holds that $ Y_r\cdot Y_c \quad \geqslant\quad |B|$. Without loss of generality, assume
that $Y_r \geqslant Y_c$. Then $Y_r^2 \geqslant |B|$ and thus $Y_r \geqslant \sqrt{|B|}$. Since event
$\mathcal{B}$ holds and every \emph{gray} row contains a \emph{non-black} cell adjacent to a \emph{black}
one, it holds that
$$
|N(I)| \, \geqslant \, Y_r \frac{R^2}{\lambda} \, \geqslant \, \sqrt{|B|}\frac{R^2}{\lambda}
$$
By using again the fact (implied by event $\mathcal{B}$) that every cell contains at most $\lambda R^2$ nodes, we have that $|B|\lambda R^2 \geqslant |I|$ and thus $\sqrt{|B|} \geqslant \frac{\sqrt{|I|}}{\sqrt{\lambda}R}$. It follows that
$$
|N(I)| \, \geqslant \, \frac{R\sqrt{|I|}}{\lambda\sqrt{\lambda}} \, \geqslant \, \beta R\sqrt{|I|}
$$
and the expansion property holds.
\qed
\end{quote}

\noindent Since, by Claim~\ref{claim:density}, event $\mathcal{B}$ occurs with probability at least $1 - \frac{1}{n^2}$, Claims~\ref{claim:exp1} and \ref{claim:exp2} imply that also the expansion properties will hold with probability $1 - \frac{1}{n^2}$.
\qed

\noindent Thanks to the general bound given by Corollary~\ref{cor:big}, the above expansion properties can be exploited in order to bound the flooding time in stationary geometric-MEG.

\begin{theorem}\label{thm:geoflooding}
Let $\mathcal{G}(n, r, R, \epsilon)$ be a stationary geometric-MEG. If $r \geqslant 0$ and $c\sqrt{\log n} \leqslant R \leqslant \sqrt{n}$ for a sufficiently large constant $c$, then the flooding time of $\mathcal{G}(n, r, R, \epsilon)$ is w.h.p.
$$
\mathcal{O}\left( \frac{\sqrt{n}}{R} + \log\log R \right)
$$
\end{theorem}
\proof From Theorem~\ref{thm:geomexpan}, the stationary geometric-MEG $\mathcal{G}(n, r, R, \epsilon)$
enjoys, with probability $1 - \frac{1}{n^2}$, the following expansion properties:
\begin{itemize}
\item $(h, \alpha\frac{R^2}{h})$-expander for $1 \leqslant h \leqslant \alpha R^2$
\item $(h, \beta\frac{R}{\sqrt{h}})$-expander for $\alpha R^2 \leqslant h \leqslant n/2$.
\end{itemize}
Thus, by applying Corollary~\ref{cor:big}, we obtain that flooding time is w.h.p.
$$
\mathcal{O}\left( \sum_{h = 1}^{\alpha R^2} \frac{1}{h\log(1 + \alpha\frac{R^2}{h})}  + \sum_{h =
\alpha R^2}^{n/2} \frac{1}{h\log(1 + \beta\frac{R}{\sqrt{h}})}\right)
$$

\noindent We now evaluate the above two sums separately. For the sake of convenience, set $T = \alpha R^2$. It
holds that
$$
\sum_{h = 1}^{T} \frac{1}{h\log(1 + \frac{T}{h})} \ \leqslant\ 2\sum_{h = 1}^{T} \frac{1}{h\log(1 + \frac{T}{h})} \frac{T}{(T + h)}
$$

\noindent This holds since $\frac{T}{(T + h)} \geqslant 1/2$ for $h \leqslant T$. Moreover,

\begin{eqnarray*}
\sum_{h = 1}^{T} \frac{1}{h\log(1 + \frac{T}{h})} \frac{T}{(T + h)}
& = & \frac{T}{(T + 1)\log(T + 1)} + \sum_{h = 2}^{T} \frac{1}{h\log(1 + \frac{T}{h})} \frac{T}{(T + h)} \\
& \leqslant & 1 + \int_1^T \frac{T}{x(T + x)\log(1 + \frac{T}{x})}dx \\
& = & 1 + [ -\log\log (1 + \frac{T}{x})]_1^T = \log\log(T) + c
\end{eqnarray*}

\noindent where $c$ is a constant. Therefore we have shown that
$$
\sum_{h = 1}^{\alpha R^2} \frac{1}{h\log(1 + \alpha\frac{R^2}{h})}  \  = \  \mathcal{O}(\log\log R)
$$

\noindent Now consider the second sum. By using the inequality $\log(1 + x) \geqslant \frac{x}{1 + x}$ we have
that
$$ \sum_{h = \alpha R^2}^{n/2} \frac{1}{h\log(1 + \beta\frac{R}{\sqrt{h}})}
\,\leqslant\, \sum_{h = \alpha R^2}^{n/2} \frac{\sqrt{h} + \beta R}{h\beta R}
\,\leqslant\, \frac{1 + \frac{\beta}{\sqrt{\alpha}}}{\beta R}\sum_{h = \alpha R^2}^{n/2} \frac{1}{\sqrt{h}}.
$$
\noindent where the last inequality comes from inequality
$$
\sqrt{h} + \beta R \leqslant (1 + \frac{\beta}{\sqrt{\alpha}}) \sqrt{h}
$$
for $h \geqslant \alpha R^2$. Moreover, it holds that
$$
\sum_{h = \alpha R^2}^{n/2} \frac{1}{\sqrt{h}} \ \leqslant\  \int_{\alpha R^2 - 1}^{n/2}
\frac{dx}{\sqrt{x}} \ \leqslant \  2\sqrt{n}
$$

\noindent By combining the above inequalities we obtain
$$
\sum_{h = \alpha R^2}^{n/2} \frac{1}{h\log(1 + \beta\frac{R}{\sqrt{h}})} \ \leqslant\  2\frac{1 +
\frac{\beta}{\sqrt{\alpha}}}{\beta R}\sqrt{n}
$$

\noindent that is,
$$
\sum_{h = \alpha R^2}^{n/2} \frac{1}{h\log(1 + \beta\frac{R}{\sqrt{h}})} \ = \ \mathcal{O}\left(\frac{\sqrt{n}}{R} \right)
$$
\qed

\noindent We remark that the proof of the expansion properties of Theorem~\ref{thm:geomexpan} only relies on the fact that the stationary distribution of node positions is almost uniform. In fact we can get the same expansion properties for any mobility model yielding a stationary distribution of node position that is uniform or almost uniform. As mentioned in the Introduction, several relevant mobility models enjoy this \emph{uniformity} property. So, thanks to our Theorem~\ref{theorem:big}, we can get an upper bound on flooding time similar to that of Theorem~\ref{thm:geoflooding}.

\noindent Next theorem shows a lower bound on flooding time in stationary geometric-MEG.

\begin{theorem}\label{thm:geolb}
Let $\mathcal{G}(n, r, R, \epsilon)$ be a stationary geometric-MEG. If $r\geqslant 0$, then its 
flooding time is w.h.p.
$$
\Omega\left(\frac{\sqrt{n}}{R + r}\right)
$$
\end{theorem}
\proof Since the geometric-MEG is stationary, it is not hard to see that, w.h.p., at time 0 there exist at least two nodes $u$ and $v$ that are at distance greater than $\sqrt{n}/2$. Consider the flooding process with source node $v$. Let $\mathbf{x}_0$ be the position of $v$ at time 0. For any $t$, let $d_t$ be the minimum distance from $\mathbf{x}_0$ that node $u$ has ever reached during the first $t$ time steps. It is immediate to see that $d_{t+1} \geqslant d_t - r$. Since $d_0 \geqslant \sqrt{n}/2$, it holds that $d_t \geqslant \sqrt{n}/2 - r\cdot t$.

\noindent Let $D_t$ be the maximal distance from  $\mathbf{x}_0$  that any informed node has ever reached during the first $t$ time steps. It is easy to see that $D_{t+1} \leqslant D_t + R + r$. Since $D_0 = 0$, it holds that $D_t \leqslant (R + r)t$.

\noindent Let $\tau$ be the time step in which node $u$ gets informed. It must be the case that $D_\tau \geqslant d_\tau$. It follows that
$$
(R + r)\tau \ \geqslant\  D_\tau \ \geqslant\  d_\tau\ \geqslant\  \sqrt{n}/2 - r\cdot \tau.
$$
It follows that $\tau \geqslant  \sqrt{n}/ (2(R + 2r))$. Therefore, the flooding cannot be completed
in less than $\Omega\left(\frac{\sqrt{n}}{R + r}\right)$ time steps.
\qed

\noindent By comparing Theorem~\ref{thm:geoflooding} and Theorem~\ref{thm:geolb} we obtain the following

\begin{cor}\label{cor:geotight}
Let $\mathcal{G}(n, r, R, \epsilon)$ be a stationary geometric-MEG. If  $r =
\mathcal{O}(R)$, and $c\sqrt{\log n} \leqslant R \leqslant \frac{\sqrt{n}}{\log\log n}$ for a
sufficiently large constant $c$, then the  flooding time of $\mathcal{G}(n, r, R, \epsilon)$ is w.h.p.
$$
\Theta\left(\frac{\sqrt{n}}{R}\right)
$$
\end{cor}

\noindent Under the very reasonable conditions of the above corollary, the general bound on flooding time in Markovian evolving graphs thus turns out to be asymptotically tight for stationary geometric-MEG.

\section{Edge-Markovian evolving graphs}\label{sec::edge}
We recall the model introduced in \cite{CMMPS08,W80}. An edge-MEG $\mathcal{M}(n,p,q)=  \{G_t  \;:\;
t \in \mathbb{N} \}$ is a Markov chain such that  $ G_t = ([n],E_t)$ with
\[
E_t =  \left\{ e \in \binom{[n]}{2} \;:\; X_t(e) = 1 \right\}
\]
where    $\{X_t(e) \;:\; e \in \binom{[n]}{2} \}$ are independent Markov chains  with transition matrix
\[
M = \left(
\begin{array}{c|cc}
& 0 & 1 \\[1mm]
\hline
0 & 1 - p & p \\[2mm]
1 & q & 1 - q
\end{array}
\right)
\]

\noindent Remind that $p$ is the birth-rate and $q$ is the death-rate and notice that an edge-MEG is a Markovian evolving  graph according to Definition \ref{def::Markovevolv}. Observe that if $0 < p,q, < 1$ the Markov chains $\{X_t(e) \;:\;  t \in \mathbb{N} \}$ are irreducible and aperiodic; so there is a unique stationary distribution
\[
\pi_e=\left(\frac{q}{p+q},\frac{p}{p+q}\right)
\]
Hence, the stationary distribution of $\mathcal{M}(n,p,q)$ is $G_{n,\hat{p}}$ (i.e. Erd\"os-R\'enyi distribution in which each possible edge occurs independently with probability $\hat{p}$) where here and in the sequel
$$
\hat{p}=\frac{p}{p+q}
$$

\noindent Stationary edge-MEG enjoy the following node-expansion properties.

\begin{theorem}\label{Marco} Let  $\mathcal{M}(n,p,q)$  be an edge-MEG such that $ \hat{p} \geqslant c \frac{\log n}{n} $ for a sufficiently large constant $c$. Then, the stationary distribution of $\mathcal{M}(n,p,q)$  yields, with probability at least $1-\frac{1}{n^2}$, a $\left(h, \frac{n\hat{p}}{c}\right)$-expander for  $1\leqslant h\leqslant \frac{1}{\hat{p}}$ and a $\left(h,\frac{n}{ch}\right)$-expander for  $\frac{1}{\hat{p}}\leqslant h\leqslant \frac{n}{2}$.
\end{theorem}

\noindent The proof of the  theorem is a simple consequence of the expansion properties of the $G_{n,p}$ model (see for instance \cite{CF07}). We here provide a detailed proof for our specific setting.

\begin{lemma}\label{lvicinignp}
Let $\hat{p} \geqslant c \frac{\log n}{n}$ for a sufficiently large constant $c$. With probability $1-\frac{1}{n^4}$ for $G_{n,\hat{p}}$ it holds that for any $I\subseteq [n]$ with $|I| \leqslant
\frac{n}{2}$,
$$
|N(I)|\geqslant \min\left\{\frac{|I|n\hat{p}}{c}, \frac{n}{c}\right\}
$$
\end{lemma}
\proof
We first consider the case when $|I|\leqslant \frac{1}{\hat{p}}$ and prove that, with probability at least $1-\frac{1}{n^2}$, it holds  $|N(I)|\geqslant \frac{|I|n\hat{p}}{c}$. Then we consider the case $\frac{1}{\hat{p}}\leqslant |I|\leqslant \frac{n}{2}$ and prove that, with probability at least $1-\frac{1}{n^2}$, it holds $|N(I)|\geqslant \frac{n}{c}$.

\noindent Let $m=|I|\leqslant \frac{1}{\hat{p}}$. For any $u\in[n]\setminus I$ consider the random variable $X_u$ so that $X_u=1$ if $u\in N(I)$ and $X_u=0$ otherwise. Since $\Prob{X_v = 1} \geqslant m\hat{p}$ we have
$$
\Expec{|N(I)|}  =  \sum_{u \in[n] \setminus I} \Expec{X_u}
= (n-m)m\hat{p}
\geqslant \left( n - \frac{1}{\hat{p}} \right) m \hat{p}
\geqslant \frac{1}{2} n m \hat{p}
$$

\noindent From Chernoff's bound we get
$$
\Prob{|N(I)| \leqslant \frac{1}{c} nm\hat{p} } \leqslant e^{-\frac{1}{4}
nm\hat{p}\left(\frac{c-2}{c}\right)^2}
\leqslant  e^{-\frac{1}{4} m \log n\frac{(c-2)^2}{c}}
\leqslant n^{-\frac{c-4}{4} m}
$$

\noindent Therefore
\begin{eqnarray*}
\Prob{\exists I \subseteq [n], \, 1 \leqslant |I| \leqslant 1/\hat{p} \;:\; |N(I)| \leqslant \frac{1}{c} |I|\hat{p}}
& \leqslant & \sum_{\begin{array}{c} I \subseteq [n] \\ 1 \leqslant |I| \leqslant 1/\hat{p} \end{array}} \Prob{|N(I)| \leqslant \frac{1}{c} n|I|\hat{p}} \\
& \leqslant &  \sum_{m=1}^{\lfloor 1/\hat{p} \rfloor} \binom{n}{m} n^{-\frac{c-4}{4} m}\\
& \leqslant & \sum_{m=1}^{\lfloor 1/\hat{p} \rfloor} n^m n^{-\frac{c-4}{4} m} \leqslant \frac{1}{\hat{p}} n^{-\frac{c-8}{4} } \leqslant n^{-\frac{c-12}{4}}
\end{eqnarray*}

\noindent And the thesis follows if we choose $c \geqslant 20$.

\smallskip
\noindent Now consider the case where  $\frac{1}{\hat{p}}\leqslant |I|=m\leqslant \frac{n}{2}$. Notice that $|N(I)|\leqslant \frac{n}{c}$ if and only if there exists a set $A\subseteq [n]\setminus (I\cup N(I))$ such that $|A|\geqslant n-m-\frac{n}{c}$. Hence
$$
\Prob{\exists I \subseteq [n],  |I|=m: |N(I)| \leqslant \frac{n}{c}  } = \binom{n}{m}
\binom{n-m}{|A|}(1-\hat{p})^{m|A|}
$$

\noindent From the following inequalities
\begin{itemize}

\item $\binom{n}{m} \leqslant n^m \leqslant e^{\frac{mn\hat{p}}{c}} $

\item $\binom{n-m}{|A|} = \binom{n-m}{n-m-|A|} \leqslant \binom{n-m}{\frac{n}{c}}\leqslant
\binom{n}{\frac{n}{c}}\leqslant (ec)^{\frac{n}{c}}= e^{\frac{n}{c}\log (ec)}\leqslant
e^{mn\hat{p}\left(\frac{1}{c}+\frac{\log c}{c}\right)}$

\item $(1-\hat{p})^{m|A|}\leqslant e^{-m\hat{p}|A|}\leqslant e^{-m\hat{p}\left(n-\frac{n}{2}-\frac{n}{c}\right)}=e^{-mn\hat{p}\left(\frac{1}{2}-\frac{1}{c}\right)}$
\end{itemize}
by choosing $c$ sufficiently large, we get
$$
\Prob{\exists I \subseteq [n], \; |I|=m: |N(I)| \leqslant \frac{n}{c}  } \leqslant
e^{-mn\hat{p}\left(\frac{1}{2}-\frac{3}{c}-\frac{\log c}{c}\right)}
\leqslant
e^{-\frac{n}{5}}
$$
Hence
\begin{eqnarray*}
\Prob{\exists I \subseteq [n],  \frac{1}{\hat{p}} \leqslant |I|  \leqslant \frac{n}{2} : |N(I)|
\leqslant \frac{n}{c} }
& \leqslant & \sum_{m=\lfloor 1/\hat{p} \rfloor}^{\lceil n/2 \rceil} \Prob{\exists I \subseteq [n], |I|=m: |N(I)| \leqslant \frac{n}{c}} \\
& \leqslant & \sum_{m=\lfloor 1/\hat{p} \rfloor}^{\lceil n/2 \rceil}  e^{-\frac{n}{5} } \leqslant ne^{-\frac{n}{5} } \leqslant n^{-2}
\end{eqnarray*}
where the last inequality holds for sufficiently large $n$.
\qed

\noindent The expansion properties of stationary edge-MEG, stated in Theorem~\ref{Marco}, allow us to apply Corollary~\ref{cor:big} and thus get the following

\begin{theorem}\label{theorem:ubedge} Let $\mathcal{M}(n,p,q)$ be a stationary edge-MEG such that $\hat{p} \geqslant c \frac{\log n}{n} $ for a sufficiently large constant $c$. Then flooding time in $\mathcal{M}(n,p,q)$ is w.h.p.
$$
\mathcal{O}\left(\frac{\log n}{\log(n\hat{p})}  + \log \log (n\hat{p})\right)
$$
\end{theorem}
\proof Thanks to Theorem~\ref{Marco}, we can apply Corollary \ref{cor:big} with sequence
$$
k_i=\left\{\begin{array}{l}
\frac{n\hat{p}}{c} \quad \mbox{ for $1\leqslant i\leqslant \left\lfloor \frac{1}{\hat{p}}\right\rfloor$}\\
\frac{n}{ci}\quad \mbox { for $\left\lfloor\frac{1}{\hat{p}}\right\rfloor < i\leqslant  \frac{n}{2}$}\\
\end{array}
\right.
$$
Thus we obtain that the order of flooding time is w.h.p. bounded by
$$
\sum_{i=1}^{\lfloor 1/\hat{p}\rfloor} \frac{1}{i\log (1 + \frac{n\hat{p}}{c})}+
\sum_{i=\lfloor 1/\hat{p}\rfloor+1}^{\lceil n/c\rceil-1} \frac{1}{i\log (1 + \frac{n}{ci})}
+ \sum_{i=\lceil n/c\rceil}^{n/2} \frac{1}{i\log (1 + \frac{n}{ci})}
$$

\noindent We now evaluate the above sums separately. For the first sum, by using
$\sum_{i=1}^{m}\frac{1}{i}\leqslant \log m+1$, we have
$$
\sum_{i=1}^{\lfloor 1/\hat{p}\rfloor} \frac{1}{i\log (1 + \frac{n\hat{p}}{c})}=
\frac{\log\frac{1}{\hat{p}}+1}{\log (1 + \frac{n\hat{p}}{c})}= \mathcal{O}\left(\frac{\log
n}{\log(n\hat{p})} \right)
$$

\noindent For the second sum, by using  $\log (1+x) \geqslant \log x$, we have
\begin{eqnarray*}
\sum_{i=\lfloor 1/\hat{p}\rfloor+1}^{\lceil n/c\rceil-1} \frac{1}{i\log (1 + \frac{n}{ci})}
& \leqslant & \sum_{i=\lfloor 1/\hat{p}\rfloor+1}^{\lceil n/c\rceil-1}\frac{1}{i\log \frac{n}{ci}}
\leqslant \int_{\lfloor1/\hat{p}\rfloor}^{\lceil n/c\rceil-1} \frac{1}{x\log \frac{n}{cx}}dx \\
& = & \left[-\log\log\frac{n}{cx}\right]_{\lfloor1/\hat{p}\rfloor}^{\lceil n/c\rceil-1} = \mathcal{O}(\log\log ( n\hat{p}))
\end{eqnarray*}

\noindent For the third sum, we apply $\log (1+x)\geqslant x/(1+x)$ for $x<1$ and get
$$
\sum_{i=\lceil n/c\rceil}^{n/2} \frac{1}{i\log (1 + \frac{n}{ci})}
\, \leqslant \, \sum_{i=\lceil n/c\rceil}^{n/2}\frac{\left(1+\frac{n}{ci}\right)}{i\frac{n}{ci}}
\, = \, \sum_{i=\lceil n/c\rceil}^{n/2} \left(\frac{c}{n}+\frac{1}{i}\right)
\, \leqslant \, \sum_{i=\lceil n/c\rceil}^{n/2} \left(\frac{c}{n}+\frac{c}{n}\right)
\, = \, \mathcal{O}(1)
$$
\qed

\noindent Next theorem gives a lower bound on flooding time in stationary edge-MEG.
\begin{theorem}\label{theorem:edgelower}
Let $\mathcal{M}(n,p,q)$ be a stationary edge-MEG such that $\hat{p} \geqslant c \frac{\log n}{n}$ for
a sufficiently large constant $c$. Then the flooding time of $\mathcal{M}(n,p,q)$ is w.h.p.
$$
\Omega\left(\frac{\log n}{\log(n\hat{p})} \right)
$$
\end{theorem}

\proof Let $\Delta_t$ be the random variable indicating the maximal node degree of $G_t$. Since the marginal distribution of $G_t$ is Erd\"os-R\'enyi $G_{n, \hat{p}}$ then, for a sufficiently large $c$ (say $c = 4$), it holds that $\Prob{\Delta_t > 2 n \hat{p}} < 1/n^2$. By applying the union bound the probability that a time step $t < n$ exists such that $\Delta_t > 2 n \hat{p}$ is at most $1/n$. Thus, the number of informed nodes at time step $t < n$ is at most $(2n\hat{p})^t$ w.h.p., and this number is less than $n/2$ for $t < \frac{\log (n/2)}{\log (2n\hat{p})}$.
\qed

\noindent By comparing the upper bound of Theorem~\ref{theorem:ubedge} and the lower bound of
Theorem~\ref{theorem:edgelower} we obtain the following

\begin{cor} Let $\mathcal{M}(n,p,q)$  be a stationary edge-MEG such that $c \frac{\log n}{n} \leqslant \hat{p} \leqslant \frac{n^{\frac{1}{\log\log n}}}{n}$, for a sufficiently large constant $c$. Then flooding time in $\mathcal{M}(n,p,q)$ is w.h.p.
$$
\Theta\left(\frac{\log n}{\log(n\hat{p})} \right)
$$
\end{cor}

\section{Conclusions} \label{sec::concl}
We showed that in geometric-MEG, under some conditions on the maximal node speed and transmission radius, node mobility has an almost negligible impact on flooding time: the latter turns out to be equivalent to the diameter of the static stationary graph. A similar fact holds for edge-MEG as well.

\noindent After the conference version of this paper, an improved, \emph{dynamic} version of our expansion technique has been derived in \cite{CPS09} in order to obtain almost tight bounds for the flooding time of highly-sparse and disconnected geometric-MEG when the maximal node speed is larger than the transmission radius. So, in this case, mobility significantly speeds-up flooding time with respect to the static case.

An important open issue is to provide analytical bounds for the flooding time of evolving graphs that are somewhat
\emph{non homogeneous}. Interesting examples are the evolving graphs yielded by node performing random walks over highly-irregular support graphs and those yielded by nodes moving according the random-waypoint model over a non-convex, irregular region.

\bibliographystyle{plain}
\bibliography{mrn}

\end{document}